\documentclass{article}

\usepackage[nonatbib,preprint]{neurips_2021}

\usepackage[utf8]{inputenc} 
\usepackage[T1]{fontenc}    
\usepackage{hyperref}       
\usepackage{url}            
\usepackage{booktabs}
\usepackage{mathtools}
\usepackage{amsfonts}       
\usepackage{nicefrac}       
\usepackage{microtype}      
\usepackage{xcolor}         
\usepackage{color,soul}

\title{MEDUSA: Multi-scale Encoder-Decoder Self-Attention Deep Neural Network Architecture for Medical Image Analysis}

\author{
  \normalfont{Hossein Aboutalebi\textsuperscript{1, 6,7},Maya Pavlova \textsuperscript{3, 7}, Hayden Gunraj \textsuperscript{3, 7}, Mohammad Javad Shafiee\textsuperscript{1, 3, 6, 7}}\\
Ali Sabri \textsuperscript{2}, Amer Alaref \textsuperscript{4,5},
 Alexander Wong\textsuperscript{1, 3, 6, 7}\\
 \textsuperscript{1}{Department of Computer Science, University of Waterloo, Canada}\\
  \textsuperscript{2}Department of Radiology, Niagara Health, McMaster University, Canada \\
    \textsuperscript{3}{Department of Systems Design Engineering, University of Waterloo, Canada}\\
  \textsuperscript{4}Department of Diagnostic Imaging, Northern Ontario School of Medicine, Canada \\
    \textsuperscript{5}Department of Diagnostic Radiology, Thunder Bay Regional Health Sciences Centre, Canada\\
 \textsuperscript{6}Waterloo AI Institute, University of Waterloo, Waterloo, Ontario, Canada\\
  \textsuperscript{7}\{haboutal, mspavlova,  hayden.gunraj, mjshafiee, a28wong\}@uwaterloo.ca \\
}

\begin{document}

	 \maketitle

\begin{abstract}
Medical image analysis continues to hold interesting challenges given the subtle characteristics of certain diseases and the significant overlap in appearance between diseases. In this work, we explore the concept of self-attention for tackling such subtleties in and between diseases.  To this end, we introduce MEDUSA, a multi-scale encoder-decoder self-attention mechanism tailored for medical image analysis.  While self-attention deep convolutional neural network architectures in existing literature center around the notion of multiple isolated lightweight attention mechanisms with limited individual capacities being incorporated at different points in the network architecture, MEDUSA takes a significant departure from this notion by possessing a single, unified self-attention mechanism with significantly higher capacity with multiple attention heads feeding into different scales in the network architecture.  To the best of the authors' knowledge, this is the first ``single body, multi-scale heads'' realization of self-attention and enables explicit global context amongst selective attention at different levels of representational abstractions while still enabling differing local attention context at individual levels of abstractions.  With MEDUSA, we obtain state-of-the-art performance on multiple challenging medical image analysis benchmarks including COVIDx, RSNA RICORD, and RSNA Pneumonia Challenge when compared to previous work. Our MEDUSA model is publicly available.

\end{abstract}

\vspace{-0.5cm}
\section{Introduction}
\vspace{-0.25cm}

The importance of medical imaging in modern healthcare has significantly increased in the past few decades and has now become integral to many different areas of the clinical workflow, ranging from screening and triaging, to diagnosis and prognosis, to treatment planning and surgical intervention.  Despite the tremendous advances in medical imaging technology, an on-going challenge faced is the scarcity of expert radiologists and the difficulties in human image interpretation that result in high inter-observer and intra-observer variability.  As a result, and due to advances in deep learning~\cite{lecun2015deep,bengio2009learning, glorot2010understanding} and especially convolutional neural networks~\cite{he2015deep, simonyan2014very,krizhevsky2012imagenet,zagoruyko2016wide}, there has been significant research focused on computer aided medical image analysis to streamline the clinical imaging workflow and support clinicians and radiologists to interpret medical imaging data more efficiently, more consistently, and more accurately.

From a machine learning perspective, the area of medical image analysis continues to hold some very interesting challenges that have yet to be solved by the research community.  There are two particularly interesting challenges worth deeper exploration when tackling the challenge of medical image analysis.  First, certain diseases have very subtle characteristics particularly at the early stages of the disease.  For example, in the case of infection due to the severe acute respiratory syndrome coronavirus 2 (SARS-CoV-2) virus, which is the cause of the on-going COVID-19 pandemic, the signs of lung infections often manifests itself at the earlier stage as faint opacities in the mid and lower lung lobes that can be difficult to characterize and distinguish from normal conditions.  Second, the visual characteristics of certain disease have high intra-disease variance as well as low inter-disease variance that makes it challenging to distinguish between diseases or characterize a given disease.  For example, many of the visual characteristics for SARS-CoV-2 infections identified in clinical literature~\cite{Wong,Warren,Toussie,Huang,Guan,zhang2021diagnosis} such as ground-glass opacities and bilateral abnormalities can not only vary significantly from patient to patient and at different stages of the disease, but are also present in other diseases such as lung infections due to bacteria and other non-SARS-CoV-2 viruses.

An interesting area to explore for tackling these two challenges found in medical image analysis in the realm of deep learning is the concept of attention~\cite{vaswani2017attention, ba2014multiple,bahdanau2014neural}.  Inspired by the notion of selective attention in human cognition where irrelevant aspects of sensory stimuli from the complex environment are tuned out in favor of focusing on specific important elements of interest to facilitate efficient perception and understanding, the concept of attention was first introduced in deep learning by Bahdanau {\it et al.} \cite{bahdanau2014neural} for the application of machine translation. The success of attention in deep learning has led to considerable breakthroughs, with the most recent being the introduction of Transformers \cite{vaswani2017attention,devlin2018bert}. Attention mechanisms in deep learning have now seen proliferation beyond natural language processing into the realms of audio perception and visual perception \cite{manchin2019reinforcement,vig2019analyzing, chorowski2015attention,woo2018cbam, wu2020visual}.

Much of seminal literature in the realm of attention for visual perception is the introduction of self-attention mechanisms within a deep convolutional neural network architecture to better capture long-range spatial and channel dependencies in visual data \cite{bello2019attention,ramachandran2019stand,woo2018cbam,hu2018squeeze}. Among the first to incorporate attention into convolutional architectures is Hu~\textit{et al.} \cite{hu2018squeeze}, who introduced channel-wise attention through lightweight gating mechanisms known as squeeze-excite modules at different stages of a convolutional neural network architecture. Woo~\textit{et al.}~\cite{woo2018cbam} extended upon this notion of light-weight gating mechanisms for self-attention through the introduction of an additional pooling-based spatial attention module which, in conjunction with the channel-wise attention module, enabled improved representational capabilities and state-of-the-art accuracy.

More recently, there has greater exploration of stand-alone attention mechanisms used both as replacement or in conjunction with convolutional primitives for visual perception. Ramachandran {\it et al.}~\cite{ramachandran2019stand} introduced a stand-alone self-attention primitive for directly replacing spatial convolutional primitives. Hu {\it et al.} \cite{hu2019local} introduced a novel local relation primitive which utilizes composability of local pixel pairs to construct an adaptive aggregation weights as a replacement for convolutional primitives.  Wu {\it et al.} \cite{wu2020visual} and Dosovitskiy  {\it et al.}\cite{dosovitskiy2020image} both studied the direct utilization of Transformer-based architectures for visual perception by tokenizing the input visual data.

A commonality between existing attention mechanisms in research literature is that selective attention is largely decoupled from a hierarchical perspective, where lightweight attention mechanisms with limited individual capacities act independently at different levels of representational abstraction.  As such, there is no direct global attentional context between scales nor long-range attentional interactions within a network architecture.  Our hypothesis is that the introduction of explicit global context amongst selective attention at different levels of representational abstractions throughout the network architecture while still enabling differing local attention context at individual levels of abstractions can lead to improved selective attention and performance.  Such global context from a hierarchical perspective can be particularly beneficial in medical image analysis for focusing attention on the subtle patterns pertaining to disease that often manifests unique multi-scale characteristics.

To test this hypothesis, we introduce MEDUSA, (\textbf{M}ulti-scale \textbf{E}ncoder-\textbf{D}ecoder \textbf{S}elf-\textbf{A}ttention), a self-attention mechanism tailored for medical image analysis.  MEDUSA takes a significant departure from existing attention mechanisms by possessing a single, unified self-attention mechanism with higher capacity and multiple heads feeding into different scales in the network architecture.  To the best of the authors' knowledge, this is the first ``single body, multi-scale heads'' realization of self-attention  where there is an explicit link between global and local attention at different scales.

The paper is organized as follows.  First, the underlying theory behind the proposed MEDUSA self-attention mechanism is explained in detail in Section 2.  The experimental results on different challenging medical image analysis benchmarks are presented in Section 3.  A discussion on the experimental results along with ablation studies are presented in Section 4.  Conclusions are drawn and future directions are discussed in Section 5.

\vspace{-0.5cm}
\section{Methodology}
\vspace{-0.3cm}
In this section, we introduce MEDUSA, a multi-scale encoder-decoder self-attention mechanism that explicitly exploits and links between both global attention and scale-specific local attention contexts through a "single body, multi-scale heads" realization to facilitate improved selection attention and performance.  First, we present the motivation behind this design.  Second, we describe the underlying theory and design of the proposed MEDUSA self-attention mechanism.  Third, we present a strategy for effectively training such  mechanism.
\vspace{-0.15cm}
\subsection{Motivation}
\vspace{-0.15cm}
While attention mechanisms have been shown in previous studies to lead to significant improvements in representational capabilities and accuracy for visual perception, their designs have involved the integration of lightweight attention blocks with limited capacity that are learnt independently in a consecutive manner. As a result, the attention blocks are largely decoupled from a hierarchical perspective, and thus there is no explicit global attention context between scales and no long-range attention interactions.  This independent attention modeling can potentially attenuate the power of attention mechanisms, especially in medical imaging data with subtle discriminative disease patterns with unique multi-scale characteristics.
As such, it is our hypothesis that the introduction of global attention context for explicitly modeling the interactions amongst selective attention at different scales alongside scale-specific local attention contexts, all learned in a unified approach can boost the representational capabilities of deep neural networks.

Motivated by that, here we learn explicit global context amongst selective attention at different levels of representational abstractions throughout the network architecture. This is achieved via a global encoder-decoder attention sub-module from the input data directly, as well as learn different local channel-wise and spatial attention contexts tailored for individual levels of abstractions via lightweight convolutional attention sub-modules. These sub-modules are connected at different layers of neural network architecture, based on both global context information from the encoder-decoder and activation response information at a given level of abstraction.  This "single body, multi-scale heads" realization of selective attention not only has the potential to improve representational capabilities but also results in efficient weight sharing through interconnections between scale-specific local attention contexts through the global attention context via the shared encoder-decoder block. This weight-sharing process allows the network to be significantly faster and allows the network to apply the same global attention map across different scales dynamically through the corresponding scale-specific attentions. As a result,  the network's training can be smoother as the attention layers are all aligned and synchronized.

In the next sections, we explain how global and local attention mechanisms are formulated in a unified structure within MEDUSA.

\vspace{-0.5cm}
\subsection{MEDUSA}
\vspace{-0.25cm}
The proposed local attentions aims to explicitly  model the global attention mechanisms in a unified framework with significantly higher capacity by incorporating multiple scale-specific heads feeding into different scales of the main network architecture. This unified framework improves the modeling capacity of the self-attention module  by feeding the local attentions with global long-range spatial context and enables them at different scales to  improve selective attention. To this end, the global modeling is formulated by an Encoder-Decoder block feeding into scale specific modules given the input sample .

Given the input sample $x\in R^{h\times w\times c}$ where $h, w, c$ are the height, width and the number of input channels, the Encoder-Decoder, $\mathcal{G}(\cdot)$, models the global attended information in a new feature space $\textbf{A}_\mathcal{G}$ with the same dimension as the input data and   $A_\mathcal{G} \in R^{h\times w\times c}$ a sample point in that space. In the next step, the vector $A_\mathcal{G}$ is fed into a  multiple scale-specific attention  attention model $\mathcal{L}(\cdot)$ to extract the consistent attended information for each scale specifically given the output feature map  $F_j$ of the scale $j$ and the global attention vector $A_\mathcal{G}$. The output of $\mathcal{L}(A_\mathcal{G},F_j)$ is final attention map corresponding to the scale $j$ combined with both global and local attentions in one unique map which is then multiplied by the feature map  at scale $j$ before feeding into the next processing block.

\begin{figure*}
\vspace{-2cm}
\centering
	\includegraphics[scale=0.5]{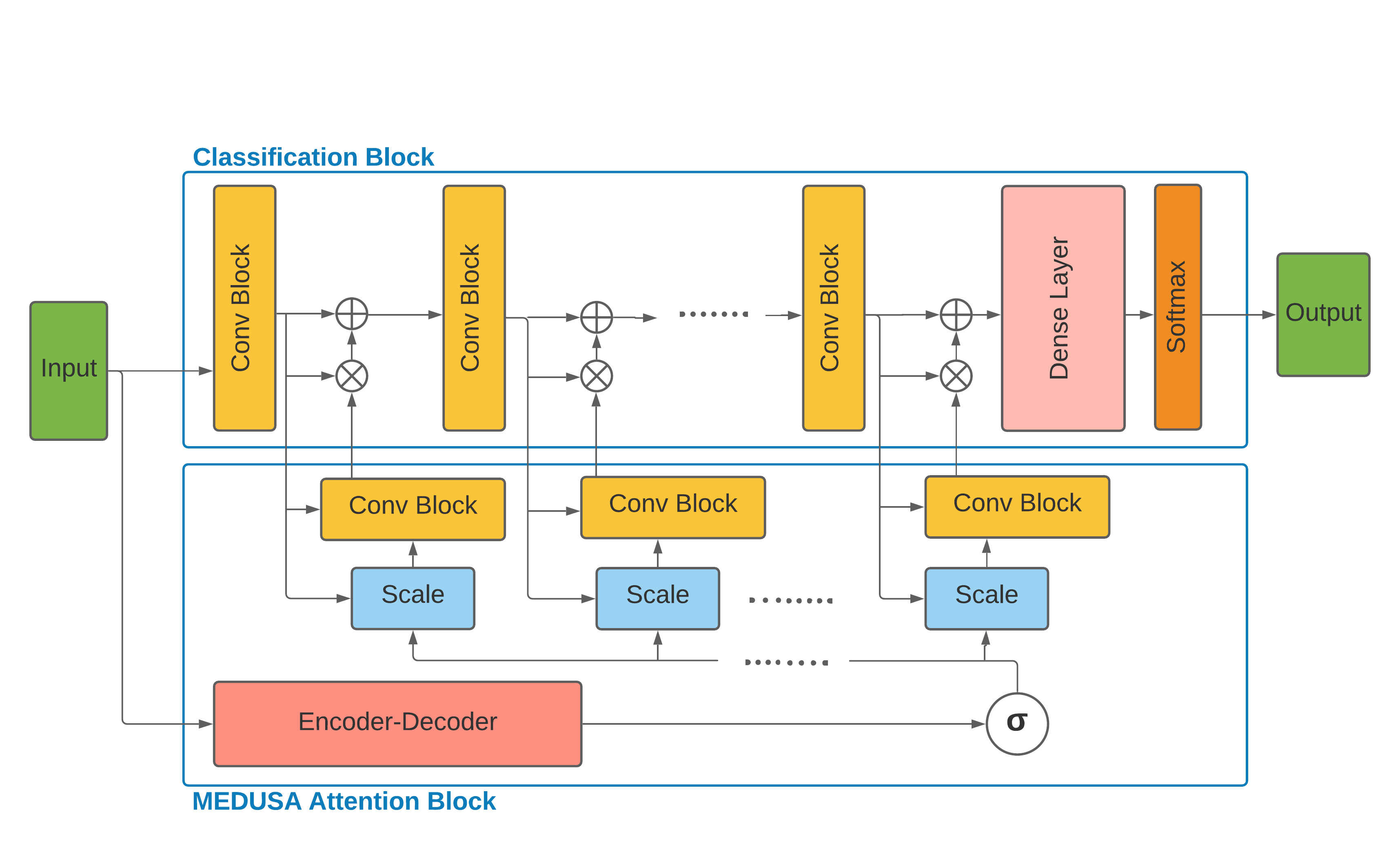}
	\caption{ Architecture of the proposed MEDUSA and how it can be incorporated to a deep neural  network. The global component of MEDUSA is fed by the input data which its output is connected to different scales through the network via the multi-scale specific module. The Scale box here refers to the bilinear interpolation operation on the output of $\sigma(A_\mathcal{G})$ fucntion based on the width and height of the feature map of the corresponding convolutional block. Here, we have only drew three convolutional blocks but the network can have an arbitrary number of convolutional blocks.}
	\label{fig:arch}
	\vspace{-0.5cm}
\end{figure*}

As seen in Figure~\ref{fig:arch}, the global attention block $\mathcal{G}(\cdot)$, generates a unified attention maps given the input sample and acts as a synchronizer among local attention blocks and interconnect them to be synced on the most important global attention information in the training process. The two main components of MEDUSA are described as follow.

\begin{itemize}
\item{\bf Global Attention: }
The Encoder-Decoder  $\mathcal{G}(\cdot)$ assimilates the global attention from the input image, with one main constraint to be consistent with different scale attentions blocks in the network. This is enforced during training when different scale attention errors are back-propagated through Encoder-Decoder  $\mathcal{G}(\cdot)$.

 The Encoder block takes the input $x$ and maps it to a latent space feature map $z\in R^{h'\times w'\times c'}$ with a downsampling network $D$. The latent space feature map $z$ has lower dimensions compared to $x$ due to the removal of  non-relevant features by the downsampling network $D$. Next, given the context vector $z$, the decoder block generates the  output $A_{\mathcal{G}}$ with the upsampling network $U$. Here it is assumed that $A_\mathcal{G}\in R^{n\times m\times c}$ which provides a weight map corresponding to every pixel of the input image. One benefit of this approach is that the weight map can provide good insights (e.g., to radiologists given the medical application) to determine and illustrate how it comes to a decision. The main purpose of using the Decoder network is to provide such human readable visualization.

\item{\bf Local Attention: }
The global attention maps need to be transformed to scale-specific attention feature map before feeding to the main network. This  task can be carried out by the  multiple scale-specific attention to connect the MEDUSA Attention block properly to the Classification blocks at different scales. Assume the Classification block consists of $J$ convolutional blocks with corresponding feature maps $F_1,\ldots, F_j, \ldots, F_J$ where $F_j \in R^{n_j\times m_j\times c_j}$.

Given the feature map $F_j$, MEDUSA Attention block infers a 3D attention map $\bar{A}_j \in R^{h_j\times h_j\times c_j}$, which  is applied on the feature map $F_j$  to transform and generate $\bar{F}_j$. The overall process can be formulated as follow:
\begin{align}\label{att1}
\bar{F}_j= \mathcal{L}_j(F_j,A_\mathcal{G})\otimes F_j + F_j
\end{align}
Where $\otimes$ denotes element-wise multiplication and $A_\mathcal{G}$ is the global attention maps generated by Encoder-Decoder $\mathcal{G}(\cdot)$.
To make MEDUSA  an efficient operation, the scale-specific modules $\mathcal{L(\cdot)}$ are formulated as:
\begin{align}
\bar{A}_j = \mathcal{L}(F_j, A_\mathcal{G}) =  \mathcal{C}_j\left(A'_j,F_j\right) \;\; \text{s.t.} \;\;\;
  A'_j = \mathcal{B}_j\Big(\sigma \left(A_\mathcal{G}\right)\Big) \label{atten2}
\end{align}
$\sigma(\cdot)$ is a Sigmoid function applied elementwise on  the tensor $A_\mathcal{G}$. This is followed by $\mathcal{B}_j$ a bilinear interpolation operation which maps the input shape width and height to $w_j \times h_j$.  $\mathcal{C}_j$ is a convolutional block with the output shape  $(h_j\times w_j\times c_j)$.
\end{itemize}

In our experiments, we found that using one convolution layer  with filter size $c_j$ is enough to get good accuracy.
The main benefit of using bilinear interpolation is to keep the same global attention feature map across different scales which will be later fine tuned for different scale through  multiple scale-specific attention. This way, we can interpret  $\sigma(A_\mathcal{G})$ to better reflects to which regions of the image the network is paying attention. As a result, the $\sigma(A_\mathcal{G})$ can be used as a visual explanation tool to further analyze the predictions made by the network.

While here we incorporate the proposed MEDUSA in a classification task, the simplicity yet effectiveness of the proposed self-attention block makes it very easy to be integrated into different deep neural network architectural for different applications.
New network architectures with the proposed self-attention module can take advantage of different training tricks, given the global attention component of MEDUSA is decoupled from the main network. This benefits the model to be trained in an iterative manner between the main model and the self-attention blocks and as a result, speeds up the convergence of the whole model. Moreover, this setup facilitates the model to take advantage of any pre-trained model for the main task.

\vspace{-0.5cm}
\subsection{Training procedure}
\vspace{-0.2cm}
As discussed earlier, the global component of the proposed self-attention block is designed  such that to be decoupled from the main network as shown in Figure \ref{fig:arch}. The global component of MEDUSA generates a unique long-range spatial context given the input sample which  is customized by the scale-specific module to generate scaled local attentions. This decoupling approach facilitates the use of training tricks in the training of both self-attention block and the main network.  To this end, we use two main tricks to train  the proposed architecture:
\begin{itemize}
\item \textbf{Transfer Learning}: Firstly, transfer learning was used to provide a better initialization of the global component (Encoder-Decoder) in the MEDUSA attention block. In particular, for the experiment on CXR dataset, we used U-Net \cite{ronneberger2015u}  as the Encoder-Decoder that was pre-trained on a large
(non-COVID-19) dataset for lung region semantic segmentation. Using a pre-trained semantic segmentation to initialize  the Encoder-Decoder, it helps  guiding the network to pay attention to relevant pixels in the image.

\item \textbf{Alternating Training}: As mentioned before, the designed structure of the proposed MEDUSA can  decouple the global component of the self-attention block from the main network. This benefits the model to use  Alternating Training technique for training the attention block and the main network sequentially, as the second training trick. During each step of the Alternating Training,   one block (the main network architecture or MEDUSA attention block) is frozen interchangeably while the other one is being trained. This way, we ensure that these two blocks learn their relatively different but related tasks concurrently. In our experiments, we  found that by using this technique, not only the network converges to the best solution faster but it also makes the training less computationally expensive in other aspects and the memory consumption during the training decreases considerably. As such, we can use larger batch sizes which decreases the training time.

\end{itemize}
\vspace{-0.5cm}
\section{Experiment Results}

\begin{figure*}
\vspace{-2cm}
\centering
	\includegraphics[scale=0.5]{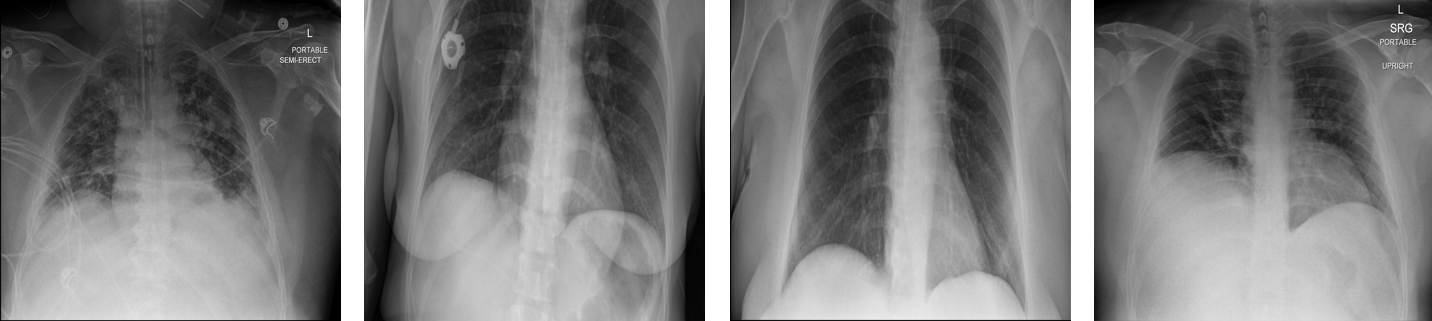}
	\caption{ Example chest X-ray images from the benchmark dataset. }
	\label{fig:sample1}
\end{figure*}
\subsection{Experimental Setup}
\vspace{-0.3cm}
In this section we describe the evaluated dataset and hyperparameters used for reporting the experimental results. \textbf{Due to the page limit, the rest of experiments are included in the supplementary.}
\vspace{-0.5cm}
\subsubsection{CXR-2 Dataset}
\vspace{-0.3cm}
To evaluate the proposed model it is trained on the largest CXR dataset consisting of 19,203 CXR images \cite{pavlova2021covid}. The dataset~\cite{pavlova2021covid} is constructed based on a cohort of 16,656 patients from at least 51 different countries. There are total of 5,210 images from 2,815 SARS-CoV-2 positive patients and the rest of images are from 13,851 SARS-CoV-2 negative patients. Interested readers can refer to \cite{pavlova2021covid} for more information on this dataset.
Figure \ref{fig:sample1} demonstrates some examples of the CXR-2 dataset.
\vspace{-0.5cm}
\subsubsection{Architecture Design}
\vspace{-0.3cm}
The global component of the proposed MEDUSA is modeled by a
 U-Net architecture~\cite{ronneberger2015u} which is used to provide the attention mechanism to a ResNet50 \cite{he2015deep} architecture for a classification task. Unlike other works in the medical imaging, we did not perform any special pre-processing on the images other than resizing the images to $480\times 480$. In our training we used the Adam optimizer~\cite{kingma2014adam} with the learning rate of 0.00008 and the batch size of 16.

\begin{table}[!t]
\caption{Sensitivity, positive predictive value (PPV), and accuracy of the proposed network (MEDUSA) on the test data from the CXR benchmark dataset in comparison to other networks.  Best results are highlighted in \textbf{bold}.}
\begin{center}
\begin{tabular}{|c|c|c|c|}
\hline
     \textbf{Architecture} & \textbf{Sensitivity ($\%$)} & \textbf{PPV ($\%$)} & \textbf{Accuracy ($\%$)} \\
\hline
ResNet-50~\cite{resnetv2} & 88.50 & 92.20 & 90.50\\
\hline
 COVID-Net~\cite{covidnet} & 93.50 & \textbf{100} & 94.00 \\
\hline
 COVID-Net CXR-2 & 95.50 & 97.00 &  96.30 \\
\hline
SE-ResNet-50 \cite{hu2018squeeze} & 90.50 & 98.90 & 94.75\\
\hline
CBAM \cite{woo2018cbam} & 70.00 & \textbf{100} & 85.00\\
\hline
MEDUSA & \textbf{97.50} & 99.00 & \textbf{98.30}\\
\hline
\end{tabular}\par
\label{tab:results1}
\end{center}
\end{table}

\begin{table}[!t]
\caption{Confusion matrix of the proposed network (MEDUSA). }
\begin{center}
\begin{tabular}{|c|c|c|}
\hline
     \textbf{SARS-CoV-2} & \textbf{Negative} & \textbf{Positive} \\
\hline
\textbf{Negative} & 198 & 2\\
\hline
\textbf{Positive} & 5 & 195\\
\hline
\end{tabular}\par
\label{tab:result2}
\end{center}
\vspace{-0.6cm}
\end{table}

\vspace{-0.4cm}
\subsection{Results \& Discussion}
\vspace{-0.2cm}
The evaluated results of the proposed method on CXR-2 dataset  are reported in Table~\ref{tab:results1}. MEDUSA provides the highest accuracy among all the other state-of-the-art techniques with at least a margin of $2\%$. Moreover, comparing to attentional based models including CBAM and SE-ResNet50 which utilize spatial attention and channel attention, MEDUSA outperform them by the accuracy margin of $3.55\%, 13.3\%$ respectively. This result illustrates the importance of formulating a unified attention model in improving the accuracy.

In addition, experimental results showed that MEDUSA and the proposed training technique leads to a much faster training convergence compared to CBAM and SE-ResNet-50 resulted in 10X speedup in convergence of the model. This shows that the unique design  proposed for MEDUSA does not impose  considerable complexity into the model's runtime cost which is a common case in other well-known attention mechanisms. Finally,  as observed in Table \ref{tab:results1}, the addition of the proposed MEDUSA to the main block of the network architecture (here the ResNet-50 for classifying COVID-19), improves the accuracy by the margin of $7.9\%$ compared to an stand alone ResNet-50, which illustrates how the proposed self-attention mechanism can help the model to better focuses on important information and leads to higher performance.
Table \ref{tab:result2} shows the confusion matrix of MEDUSA. It can be observed that the proposed attention mechanism equally increases both sensitivity and specificity of the classification model.

\vspace{-0.3cm}
\subsubsection{Global Attention}
\vspace{-0.3cm}

\begin{figure*}
    \hspace{-1.5cm}
	\includegraphics[scale=0.8]{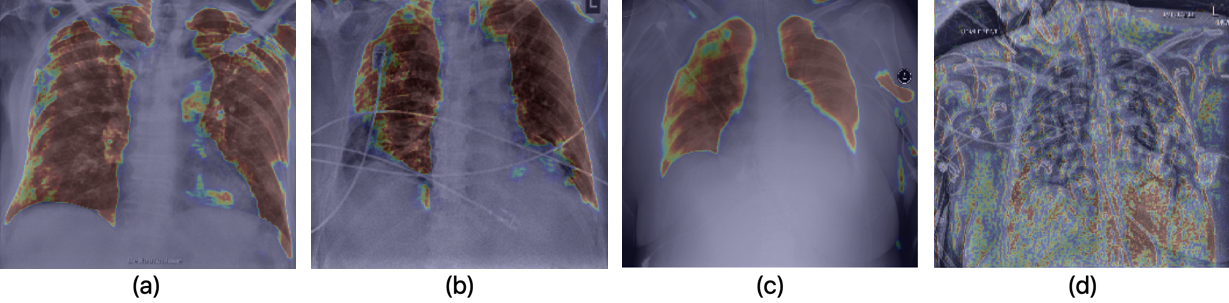}
	\caption{ Example chest X-ray images from the benchmark dataset overlaid by the global attention context. }
	\label{fig:atten1}
	\vspace{-0.3cm}
\end{figure*}

Let us now study the behaviour of the global attention sub-module of the proposed MEDUSA self-attention mechanism by visualizing its attention outputs for a variety of image examples. Figure~\ref{fig:atten1} shows the global attention outputs (i.e., the output after the $\sigma(A_{\mathcal{G}})$ function) overlaid on the input images in the form of heat maps. The red area indicates higher global attention while blue areas indicates lower global attention. The images used here is from the same CXR-2 dataset. The model correctly classifies the cases in Figure 3.a, 3.b, and 3.c while the case in Figure 3.d is incorrectly classified by the network.

It can be observed that while global attention is clearly focused on the lung region in the correctly classified cases in Figure 3.a, 3.b, and 3.c,  the global attention is not focusing on the lung region in the incorrectly classified case of Figure 3.d, which could be attributed to the poor quality of the chest CXR image compared to the correctly classified cases. This heat map visualization can help with better determining if the model is leveraging relevant information to infer the correct prediction which is very important in critical decision-making such as medical applications. In addition, it is interesting to notice that in the case of Figure 3.b, the global attention mechanism focuses away from the wires on the chest, and thus proves to be useful for avoiding attention on unrelated, non-discriminative patterns that may otherwise be leveraged for making the right decisions for the wrong reasons. Also, we can see that in Figures 3.a and 3.c,  MEDUSA helps significantly in focusing attention on the lung region for improved guidance towards relevant patterns.  We believe providing this kind of visualization can greatly enhance the trust in deep learning models especially in medical applications where the decision-making causes vital outcomes.

\vspace{-0.3cm}
\subsubsection{Local Attention }
\begin{figure*}
    \vspace{-2cm}
	\includegraphics[scale=0.6]{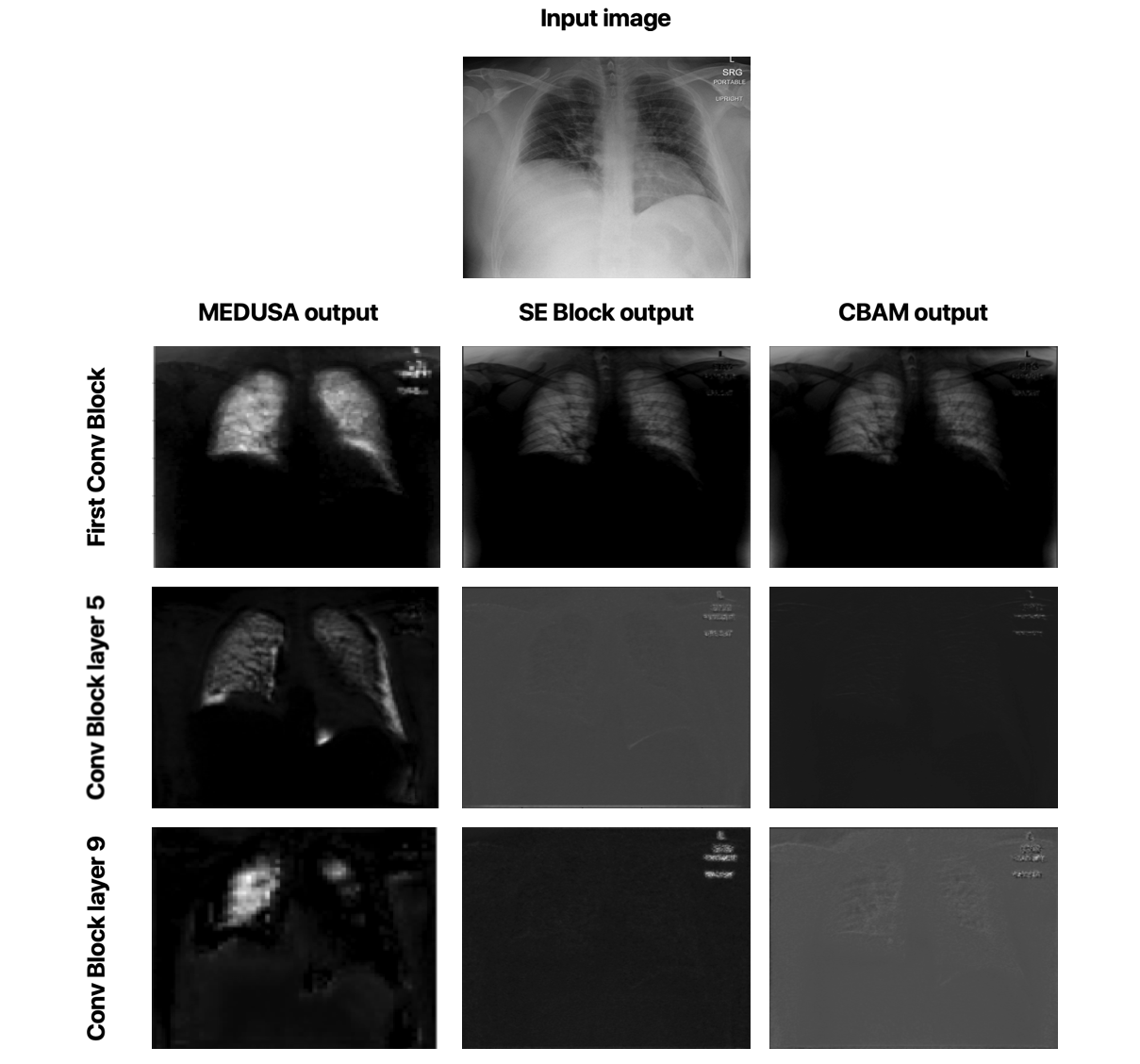}
	\caption{ Comparison between attention-enforced outputs at different convolutional blocks when using CBAM, SE Block, and MEDUSA self-attention mechanisms. Each row demonstrates the results of the attention mechanisms on a different layer of the ResNet-50 network architecture.}
	\label{fig:atten2}
	\vspace{-0.6cm}
\end{figure*}

\vspace{-0.3cm}
Let us now study the impact of the scale-specific attention sub-modules of MEDUSA towards local selective attention. Here, we compared the activation outputs of convolutional layers at different stages of the network after asserting selective attention via MEDUSA, SE, and CBAM attention mechanisms. More specifically, we study the attention-enforced outputs of three different convolutional blocks (shown in Figure \ref{fig:atten2}), to observe how the activation behaviour evolves at different stages of the network. The whiter pixels refers to higher attention-enforced activations, with all activations normalized for visualization purposes. As  seen in Figure \ref{fig:atten2}, while all tested attention mechanisms can guide attention towards relevant areas of the image (e.g., lungs) in the first convolution block, as we go to the deeper blocks, the SE and CBAM mechanisms starts to lose focus on these relevant areas. On the other hand, we can observe from the attention-enforced outputs where MEDUSA is leveraged that as we go deeper into the network, the attention-enforced activation outputs consistently focuses on the relevant areas for decision making while narrowing down focus towards more localized discriminative patterns within the broader area of interest in earlier blocks. This figure depicts the effectiveness of introducing global context alongside tailored local attention contexts at different scales, which provides a better hierarchical representation of the input image and the model can better extract higher level features that are more localized around the important pixels.

\vspace{-0.5cm}
\section{Ablation Study}
\vspace{-0.3cm}
\begin{table}[t]
\vspace{-0.05cm}
\caption{Ablation Study }
\begin{center}
\begin{tabular}{|l|c|c|c|}
\hline
     \textbf{Architecture} & \textbf{Sensitivity ($\%$)} & \textbf{PPV ($\%$)} & \textbf{Accuracy ($\%$)} \\
\hline
ResNet-50~\cite{resnetv2} & 88.50 & 92.20 & 90.50\\
\hline
 Seg Type 1  $+$ResNet-50~\cite{resnetv2} & 92.00 & 95.83 & 94.00\\
\hline
Seg Type 2  $+$ResNet-50~\cite{resnetv2} & 94.50 & 96.43 & 95.50\\
\hline
MEDUSA (Self-attention  is disabled  at test time) & \textbf{
98.50} & 88.70 & 93.00\\
\hline
MEDUSA & 97.50 & \textbf{
99.00} & \textbf{98.30}\\
\hline
\end{tabular}\par
\label{tab:ab1}
\end{center}
\vspace{-0.5cm}
\end{table}
In this section we further study the impact of MEDUSA attention block to investigate to what extent it leads the network to  improve its performance. Here  we specifically consider two different scenarios.

The first scenario is to see what happens if we turn off the MESUSA attention block on ResNet-50 after the training is done. This will show us how much impact the attention block has during the training and testing as we can compare the obtained results with base  ResNet-50 model.
In the second scenario, we answer the question of whether MEDUSA can only help the model to just ignore unnecessary information of the input sample or it provides scale specific  attentional context through the network. To  this end, the segmented result of the lung area given the input image is fed to ResNet-50. In this case, we study two types of image segmentation. In the first type, we provide the CXR image which only contains the lung region, $s(x)$ to the network, where function $s$ is the segmentation operation and $x$ is the unsegmented CXR image. In the second type of segmentation, we provide $x+s(x)$ as the input to the ResNet-50.

The results are grouped together in Table \ref{tab:ab1}. As seen, when MEDUSA is disabled at test time, while the model still retains higher accuracy than ResNet-50 baseline by $3\%$, the accuracy drops $5.3\%$ compared to the when the MEDUSA block is still enabled during the testing. This shows that not only the MEDUSA attention block causes the ResNet-50 to attain higher accuracy during the training, but it also makes the model have higher accuracy during the testing. We also observe the similar pattern when we only provide the segmented image and remove the MEDUSA block in ResNet-50. In this regard, the network loses close to $2.5\%$ accuracy on the test set. This proves that the MEDUSA attention mechanism is more than just a segmentation applied to the input  image like what has been used in the papers \cite{sarkar2021identification,khuzani2021covid}. The proposed self-attention mechanism provides multi-scale attention context which are learnt via a unified self-attention mechanism from a global context.

\vspace{-0.5cm}
\section{Conclusion}
\vspace{-0.5cm}
In this paper we proposed a novel attention mechanism so-called MEDUSA which is specifically tailored for medical imaging applications by providing a unified formulation for the attention mechanism. The  global context is modeled explicitly amongst selective attention at different scales and representational abstractions throughout the network architecture which can help to model the scale-specific attention more effectively. This unified framework provides a more coherent attention mechanism at different scales to the network leading to more accurate attention context and higher performance as a direct result.  Our results attest that the current model is not only faster than some of the predecessor but it is also able to achieve higher accuracy. While the results showed the effectiveness of the proposed attention mechanism on image based and medical applications, we aim to introduce the new version of the proposed MEDUSA in designing new architectures for other problems such as NLP and sequential data. Moreover, new training techniques to speed up the convergence and improving the model accuracy is another direction of the future work.
\section{Comprehensive Comparative Evaluation on Medical Image Analysis Datasets}
In this section, we further validate the efficacy of the proposed MEDUSA self-attention mechanism through comprehensive experiments on two popular medical image analysis datasets comparing the performance of MEDUSA to other state-of-the-art deep convolutional neural networks as well as other state-of-the-art self-attention mechanisms for deep convolutional neural networks.  First, we conducted experiments and comparative analysis on MEDUSA and other tested state-of-the-art methods on the RSNA Pneumonia Detection Challenge dataset \cite{team2018pneumonia} for the purpose of Pneumonia patient case detection. Next, we conducted experiments and comparative analysis on MEDUSA and other tested methods on a multi-national patient cohort curated by the RSNA RICORD initiative \cite{tsai2021rsna} for the purpose of severity scoring of COVID-19 positive patients.

The source  code of the proposed MEDUSA is available  \href{https://drive.google.com/file/d/1j96GxRGYMD2_00o6gHiDlUfg_m0qXwgw/view?usp=sharing}{\textbf{\underline{here}}}. All code were implemented using TensorFlow version 1.15 in Python version 3.7.

\subsection{RSNA Pneumonia Detection Challenge Dataset}

This dataset was curated by the Radiological Society of North America (RSNA) and consists of frontal-view chest x-ray images from a cohort of
26,684 patients for the purpose of Pneumonia patient case detection. The images are labeled pneumonia-positive and pneumonia-negative, with $\sim$6,000 of which being pneumonia-positive cases.  It is very important to note that, based on our deeper analysis of the data, approximately 20\% of the chest x-ray images contain significant distortions and visual anomalies. Examples of such images are shown in Figure~\ref{fig:sample1}.  Such distortions and visual anomalies make this particular dataset quite challenging and thus particularly effective for evaluating the selective attention capabilities of MEDUSA and other state-of-the-art self-attention mechanisms to focus on the right visual cues amidst such distortions.

\begin{figure*}
\centering
	\includegraphics[scale=0.28]{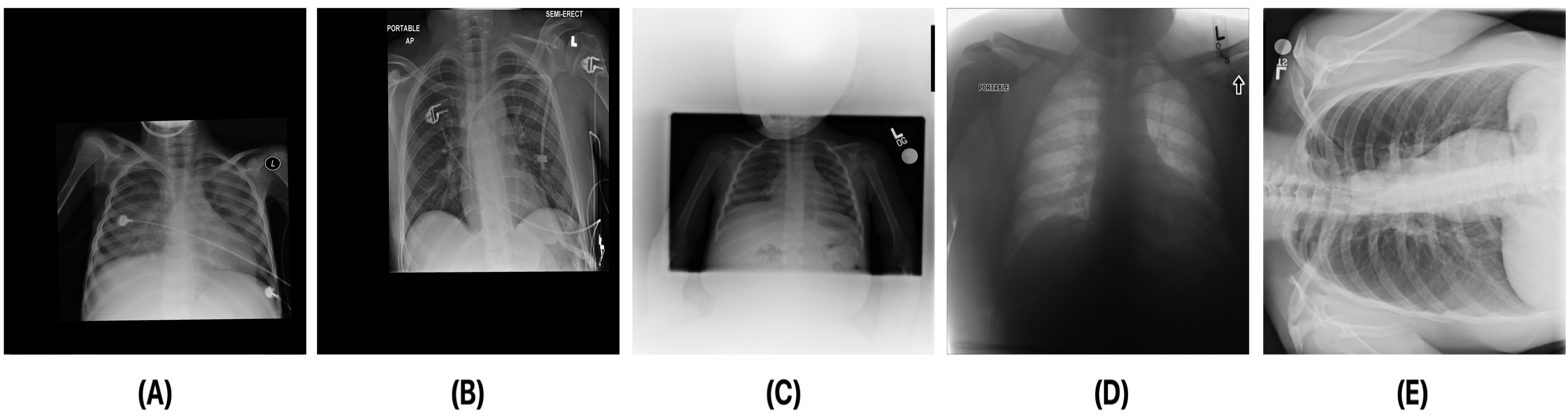}
	\caption{ Examples of chest X-ray images with different types of significant distortions and visual anomalities from the RSNA Pneumonia Challenge dataset.}
	\label{fig:sample1}
\end{figure*}

The experimental results evaluating different networks for this dataset are depicted in Table \ref{tab:results1}.  It can be observed that leveraging the proposed MEDUSA self-attention mechanism can provide significant performance improvements over other state-of-the-art self-attention mechanisms, leading to over $6\%$ higher accuracy when compared to other methods. Furthermore, leveraging MEDUSA resulted in a $14\%$ gain in sensitivity when compared to other methods. Nonetheless, CheXNet \cite{rajpurkar2017chexnet} provides a higher positive predictive value (PPV) among the tested networks at the cost of a significantly lower overall sensitivity.

\begin{figure*}

\centering
	\includegraphics[scale=0.5]{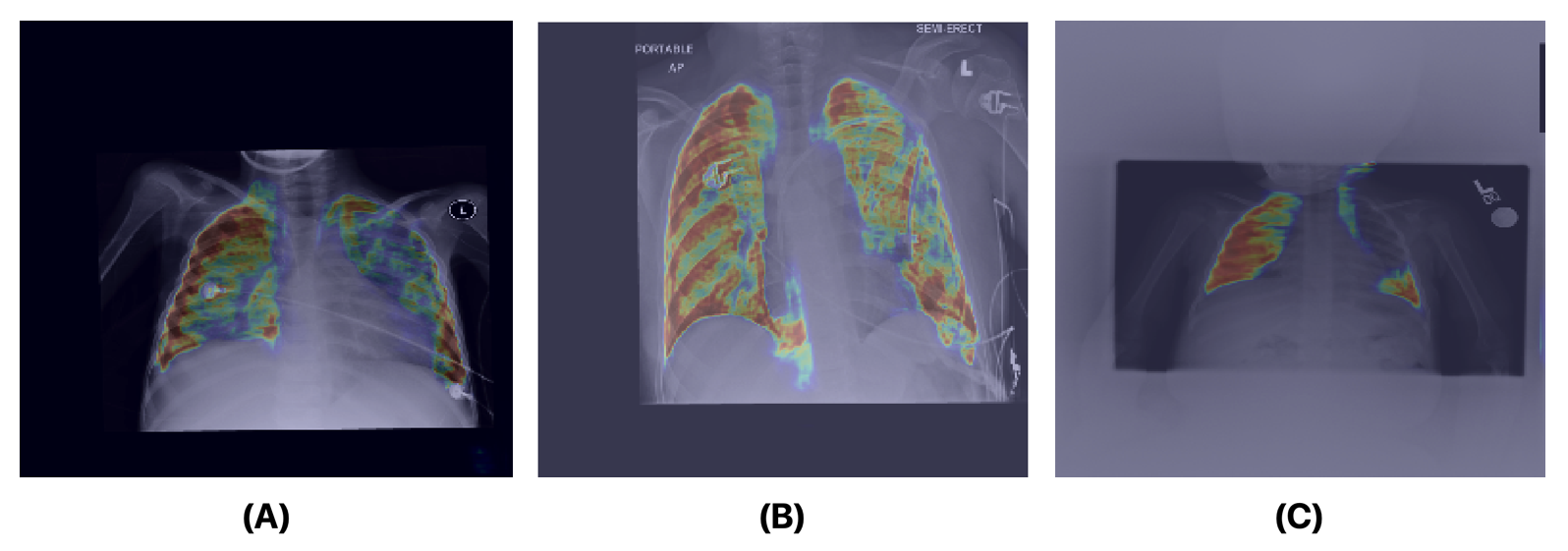}
	\caption{ Impact of MEDUSA global attention on the images with distortion. The images A, B, C are corresponding to images A, B and C in Figure~\ref{fig:sample1}.}
	\label{fig:gactivation}
\end{figure*}

\begin{figure*}

\centering
	\includegraphics[scale=0.7]{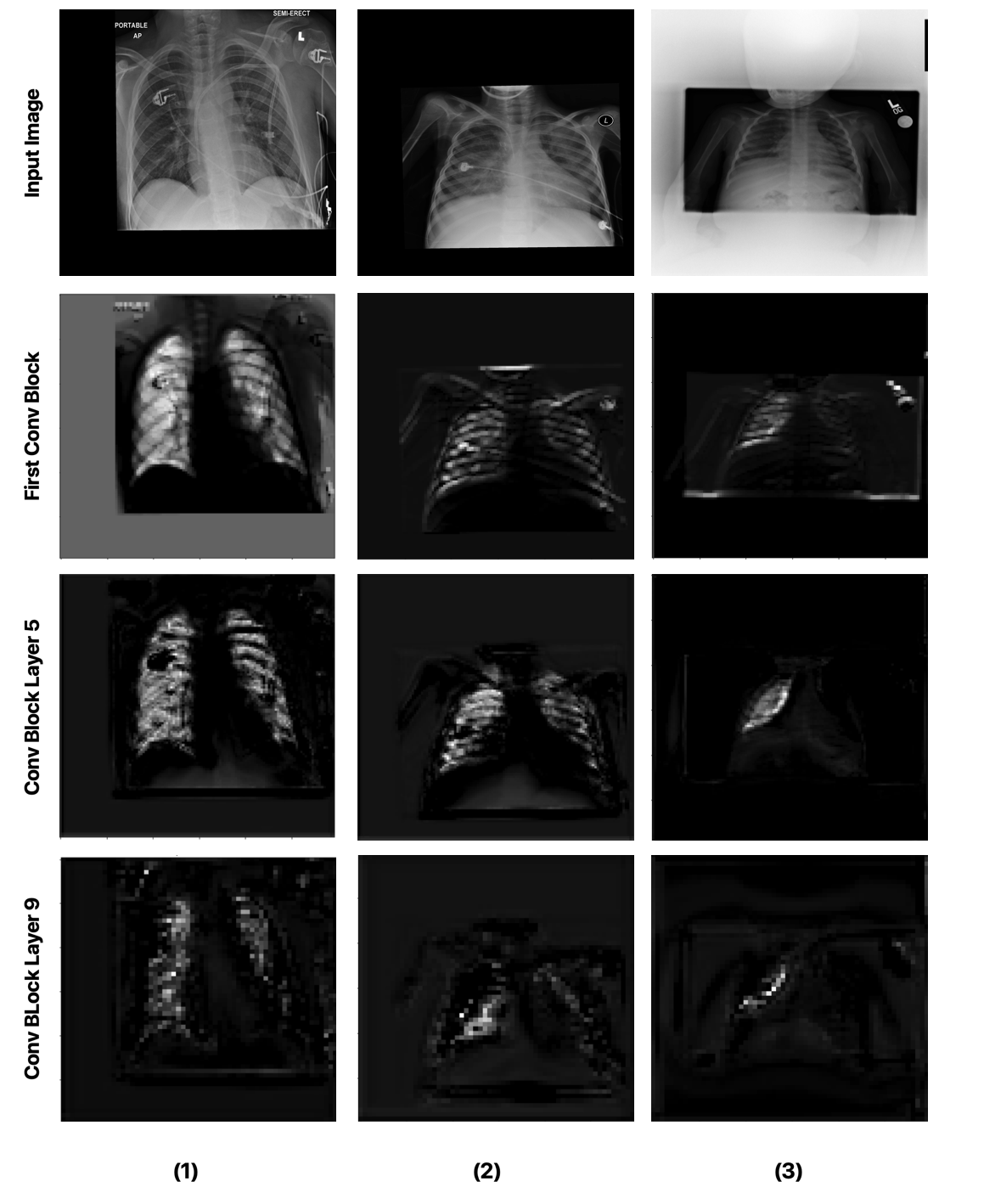}
	\caption{Impact of MEDUSA local attention on the images with distortion. Columns 1,2,3 correspond to images B, A, and C in Figure~\ref{fig:sample1}, respectively.}
	\label{fig:5}
\end{figure*}

Here, we also investigate how MEDUSA's global and local attention impact the performance of the model when the input image is distorted. Figure~\ref{fig:gactivation} and Figure~\ref{fig:5} demonstrate the global attention and local attention from MEDUSA, respectively, visualized for a subset of images showed in Figure~\ref{fig:sample1}. As seen, the global attention component  of the proposed self-attention mechanism effectively identifies the most informative regions of the image for the model to attend to, even when the images are distorted and no additional preprocessing is applied. The proposed self-attention mechanism clearly helps the model to focus on the most important information which confirmed by the quantitative results as well.

In Figure~\ref{fig:5} shows the local attention heads outputs which are adjusted based on the global attention map at different blocks in the convolutional network by the scale-specific modules of the proposed MEDUSA. As we go deeper into the network, the attention area is more localized on the relevant regions which is consistent by  the reported results in the main paper.

\subsection{ RSNA RICORD COVID-19 Severity Dataset}
\begin{figure*}
\centering
	\includegraphics[scale=0.5]{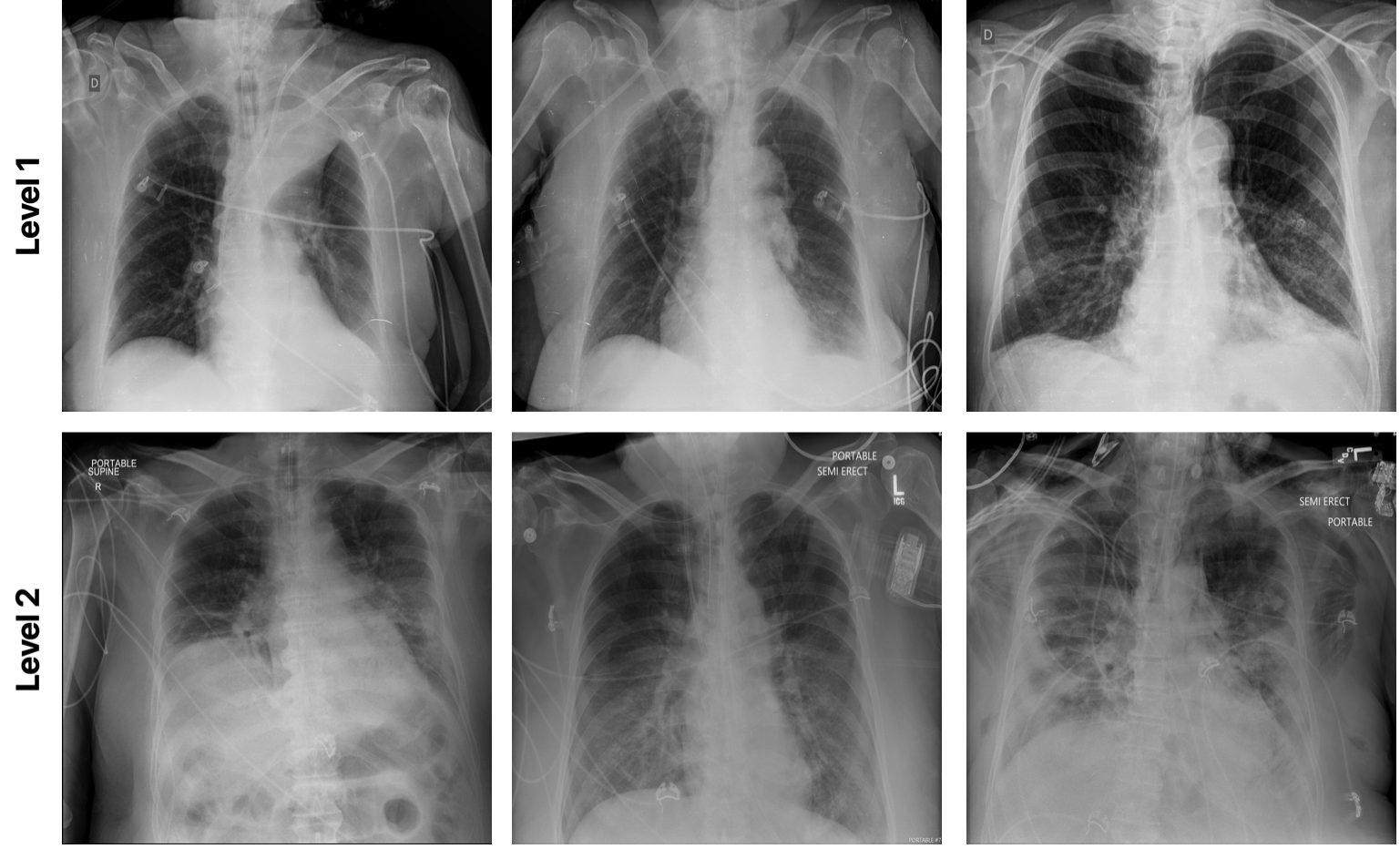}
	\caption{ Example chest X-ray images from the RSNA RICORD Covid-19 Severity dataset. }
	\label{fig:sample2}
\end{figure*}

This dataset was curated by the Radiological Society of North America (RSNA) and consists of chest x-ray images with full annotations on the severity condition score associated with COVID-19 positive patients.  Each lung is split into 3 separate zones (total of 6 zones for each patient) and the opacity is measured for each zone. Here for the experimental result the patient cases are grouped into two airspace severity levels:  Level 1: opacities in 1 or 2 zones, and Level 2: opacities in 3 or more zones. The multi-national patient cohort in this dataset consists of 909 CXR images from 258 patients.  Among the 909 CXR images, 227 images are from 129 patients with Level 1 annotation and the rest of the images are grouped with Level 2 class label.
Figure~\ref{fig:sample2} illustrates  example CXR images from this dataset for the different airspace severity level groups.

The efficacy of the proposed MEDUSA self-attention mechanism and that of other state-of-the-art methods are shown in Table~\ref{tab:results2}, with sensitivity and PPV values being reported for Level 2 patient cases. Again, we observe that MEDUSA has superior accuracy and positive predictive value (PPV) when compared to other approaches. The proposed MEDUSA self-attention mechanism provided over 8.6\% higher accuracy and over 11.2\% higher PPV than compared SE and CBAM self-attention mechanisms. While leveraging the SE self-attention mechanism resulted in the highest sensitivity in this experiment, its overall accuracy is lower due to its poor performance on Level 1 patient cases.

\begin{table}[!t]
\caption{Sensitivity, positive predictive value (PPV), and accuracy of the proposed network (MEDUSA) on the test data from the RSNA Pneumonia dataset in comparison to other networks.  Best results are highlighted in \textbf{bold}.}
\begin{center}
\begin{tabular}{|c|c|c|c|}
\hline
     \textbf{Architecture} & \textbf{Sensitivity ($\%$)} & \textbf{PPV ($\%$)} & \textbf{Accuracy ($\%$)} \\
\hline
SE-ResNet-50 \cite{hu2018squeeze} & 40.0 & 90.9 & 68.0\\
\hline
CheXNet \cite{rajpurkar2017chexnet} & 50.0 & \textbf{92.6} & 73.0\\
\hline
CBAM \cite{woo2018cbam} & 68.0 & 77.3 & 74.0\\
\hline
MEDUSA & \textbf{82.0} & 83.7 & \textbf{83.0}\\
\hline
\end{tabular}\par
\label{tab:results1}
\end{center}
\end{table}

\begin{table}[!t]
\caption{Sensitivity, positive predictive value (PPV), and accuracy of the proposed network (MEDUSA) on the test data from the RSNA RICORD Covid-19 Severity dataset in comparison to other networks.  Best results are highlighted in \textbf{bold}.}
\begin{center}
\begin{tabular}{|c|c|c|c|}
\hline
     \textbf{Architecture} & \textbf{Sensitivity ($\%$)} & \textbf{PPV ($\%$)} & \textbf{Accuracy ($\%$)} \\
\hline
SE-ResNet-50 \cite{hu2018squeeze} & \textbf{90.8} & 77.4 & 76.7\\
\hline
CheXNet \cite{rajpurkar2017chexnet} & 63.46 & 84.62 & 83.33\\
\hline
CBAM \cite{woo2018cbam} & 84.0 & 73.0 & 70.0\\
\hline
MEDUSA & 85.0 & \textbf{88.6} & \textbf{85.3}\\
\hline
\end{tabular}\par
\label{tab:results2}
\end{center}
\end{table}

\section{Ethics approval}

The study has received ethics clearance from the University of Waterloo (42235).

{\small
\bibliographystyle{IEEEtran}
 \bibliography{egbib}
}
\end{document}